\documentclass[prd,amssymb,twocolumn,10pt]{revtex4-1}
\usepackage{amsmath,amsfonts,amssymb}
\usepackage{color,verbatim,graphicx,float}

\begin{document}

\newcommand{\dR}{\mathbb R}
\newcommand{\dC}{\mathbb C}
\newcommand{\dZ}{\mathbb Z}
\newcommand{\id}{\mathbb I}
\newcommand{\dT}{\mathbb T}
\newcommand{\red}[1]{\textcolor{red}{#1}}
\title{Quantum states of the bouncing universe}

\author{Jean Pierre Gazeau$^{\dag}$, Jakub Mielczarek$^{\ddag,\sharp}$ and W{\l}odzimierz Piechocki$^{\ddag}$
\\  $^\dag$ Laboratoire APC, Univ Paris Diderot, Sorbonne Paris Cite, 75205 Paris, France \\
$^{\ddag}$ Department of Fundamental Research, National Centre for
Nuclear Research,\\ Ho{\.z}a 69, 00-681 Warsaw, Poland \\
$^\sharp$ Institute of Physics, Jagiellonian University, Reymonta 4, 30-059 Cracow, Poland}

\date{\today}

\begin{abstract}
In this paper we study quantum dynamics of the bouncing
cosmological model. We focus on the model of the flat
Friedman-Robertson-Walker universe with a free scalar field. The
bouncing behavior, which replaces classical singularity, appears
due to the modification of general relativity along the methods of
loop quantum cosmology. We show that there exist a unitary
transformation that enables to describe the system as a free
particle with  Hamiltonian equal to canonical momentum. We examine
properties of the various quantum states of the Universe: boxcar
state, standard coherent state, and soliton-like state, as well as
Schr{\"o}dinger's cat states constructed from these states.
Characteristics of the states such as quantum moments and Wigner
functions are investigated. We show that each of these states
have, for some range of parameters, a proper semiclassical limit
fulfilling the correspondence principle. Decoherence of the
superposition of two universes is described and possible
interpretations in terms of triad orientation and
Belinsky-Khalatnikov-Lifshitz conjecture are given. Some
interesting features regarding the area of the negative part of
the Wigner function have emerged.
\end{abstract}


\maketitle

\section{Introduction}

It is known that the cosmological singularity problem of the
Friedman-Robertson-Walker (FRW) universe can be resolved, in the
sense that big bang may be replaced by big bounce, by applying
loop quantum cosmology (LQC) methods. However, quantum states
specifying quantum evolution of the universe have not been
examined in a satisfactory way yet.

There exist two alternative forms of LQC: Dirac's LQC (see e.g.
\cite{Ashtekar:2011ni,Bojowald:2006da} and references therein) and
reduced phase space (RPS) LQC (see e.g.
\cite{Dzierzak:2009ip,Malkiewicz:2009qv,Mielczarek:2011mx,Mielczarek:2012qs})
and references therein). Both approaches rely on the same form of
{\it modification} of general relativity. It consists in
approximating the curvature of connection by holonomies around
small loops with non-zero size. Going with the size to zero
removes the modification. One of the main differences between
these two approaches is an {\it interpretation} of the way of
resolving the singularity. In the Dirac LQC, one argues that the
resolution is due to strong quantum effects at the Planck scale.
In the RPS LQC, one says that it is the modification of GR by 
loop deformation of the phase space that is responsible for the 
resolution of the singularity.

In what follows we apply the RPS LQC method. It consists in first
solving dynamical constraints at the classical level and then
quantizing  the resulting classical system. This approach allows
to implement quantization easily. It gives a clear picture of
quantum dynamics for any value of an evolution parameter (time),
and enables obtaining analytical results (at least for the FRW
case).

The quantum evolution across the big bounce can give an insight
into the structure of the quantum phase. The input in describing
an evolution is a self-adjoint Hamiltonian (generator of dynamics)
together with an initial state of the universe. Taking different
initial states may lead to  different quantum evolutions. Since it
is unknown which initial state is the most natural one, we examine
generic states known in quantum physics with different properties.
Comparison of obtained results with observational cosmological
data may give {\it suggestions} concerning the choice of some
realistic initial state.

The paper is organized as follows. In Sec. II the classical
dynamics of the model is examined and the canonical transformation
simplifying the dynamics is introduced. In Sec. III we make
canonical quantization of the model. We also introduce unitary map
which corresponds to the canonical transformation introduced in Sec.
II. In Sec. IV different choices of the initial quantum state are
introduced as well as all the necessary tools which will be used
to investigate their properties.  Thereafter, in Sections V, VI
and VII, detailed analysis of the boxcar state, standard coherent
state, soliton-like state as well as  Schr\"odinger cat states
constructed from these states is performed. In Sec. VIII,
decoherence of the Schr\"odinger cat state is discussed, and
possible interpretations of such a process in the cosmological
realm are given. The issue of quantum entropy is discussed Sec.
IX. It is shown that, in contrast to the results of Ref.
\cite{Mielczarek:2012qs}, entropy of squeezing is constant for the
canonically transformed system. Then, in Sec. X, ranges of the
parameters of considered states are constrained by using the
correspondence principle between quantum and classical mechanics.
In Sec. XI, we summarize our results and draw conclusions.

\section{Classical dynamics}

The flat FRW model of the universe with a free scalar field is
described by the Hamiltonian constraint \cite{Mielczarek:2011mx}
\begin{equation}
\tilde{H} = - \frac{3}{8\pi G
\gamma^2}\frac{\sin^2(\lambda \beta)}{\lambda^2}v+\frac{p^2_{\varphi}}{2v} \approx 0,
\label{HamConst}
\end{equation}
where, as in the rest of this paper, we use the units with the
speed of light in vacuum equal to one. Here, $\gamma$ is the
Barbero-Immirzi parameter, a free parameter of the theory.
However, value of this parameter is  usually fixed from
considerations of the black hole entropy.  In particular, as
derived in Ref. \cite{Meissner:2004ju}, $\gamma \approx 0.2375$.
The parameter $\lambda$ is a `discretization scale', which is
expected to be of the order of the Planck length $l_{\text{Pl}}
=\sqrt{\hslash G} \approx 1.62 \cdot 10^{-35}$ m, but in fact
should be fixed by observational data.

The variables $\beta$ and $v$, fulfilling the Poisson bracket
$\{\beta, v\} = 4 \pi G \gamma $, parametrize the gravitational
sector. In turn, the scalar field $\varphi$ and its conjugated
momenta $p_{\varphi}$ satisfy standard relation $\{\varphi,
p_{\varphi} \} =1$, and specify sources.

Hamiltonian (\ref{HamConst}) has the symmetry
\begin{equation}
\beta \rightarrow \beta+\frac{\pi}{\lambda},
\end{equation}
so the gravitational part of phase space has topology of a
cylinder, $S^1\times \mathbb{R}$. Already classically we have
bounce type solutions in the regions: $\lambda \beta  \in [0+m\pi, \pi+m\pi]$,
for $m \in \mathbb{Z}$.

The Hamiltonian constraint  (\ref{HamConst}) can be rewritten in
the form
\begin{equation}
v^2 \sin^2(\lambda \beta) = \text{const},
\label{const}
\end{equation}
where we used the fact that $p_{\varphi}=$ const for the free
field case. It turns, that the  square root of (\ref{const}) plays
a role of the {\it physical} Hamiltonian. Based on this, {\it
dynamics} of a flat FRW cosmological model, with a freescalar
field, can be described by
\begin{equation}\label{Ham}
H = p \ \sin q,
\end{equation}
where $q\in[0,\pi]$ and $p\in \mathbb{R}$ are canonical variables
satisfying the algebra $\{q,p\} = 1$. The Hamiltonian (\ref{Ham})
occurs both in the reduced phase space approach
\cite{Mielczarek:2011mx} and standard formulation of LQC
\cite{Bojowald:2007zza}. It need not be bounded from below as it
describes the entire universe, which is an isolated system. The
$p$ variable is proportional to the \emph{volume} $v$, but we
increase its range to negative values for mathematical
convenience. Physical meaning of this volume is not clear for the
flat FRW model. It corresponds to the total volume of space if
topology is compact. The possibility of positive and negative
values of $p$ may be related to two orientations of triad. Namely,
the phase space $(\beta, v)$ is only half of the original phase
space of the model since $v :=\left|\tilde{p} V_{0}^{2/3}
\right|^{3/2}$, where $\bar{p} \in \mathbb{R}$ is a variable
parametrizing densitized triad $E^{^a_i} = \tilde{p} \delta ^a_i$.
The $V_{0}$ is a fiducial cell over which the spatial integration
is performed. By allowing positive and negative values of the
variable $p$ we recover the volume of the original phase space.
The variable $q: = \lambda \beta $ is proportional to the Hubble
factor in the classical limit ($q \ll 1$).

The Hamiltonian (\ref{Ham}) generates evolution of any phase space
function $f$ according to the equation
\begin{equation}
\frac{df}{dT} = \{f, H\}.  \label{HamEq}
\end{equation}
The $T$ variable is an intrinsic time parameter related with the
value of the scalar field. The direction of time $T$, which also
occurs in \cite{Mielczarek:2011mx}, is opposite to the direction
of the coordinate time $t$. In order to fix the directions of $T$
and $t$ one has to redefine $T$ by multiplying it and the Hamiltonian
(\ref{Ham}) by minus one. But these are only technical details
devoid of deep meaning.

Applying (\ref{HamEq}) for the canonical variables we get
\begin{eqnarray}
\frac{dq}{dT} &=& \frac{\partial H}{\partial p} = \sin q,  \\
\frac{dp}{dT} &=& -\frac{\partial H}{\partial q} = - p\,\cos q .
\end{eqnarray}

\begin{figure}[ht!]
\centering
\includegraphics[width=6cm,angle=0]{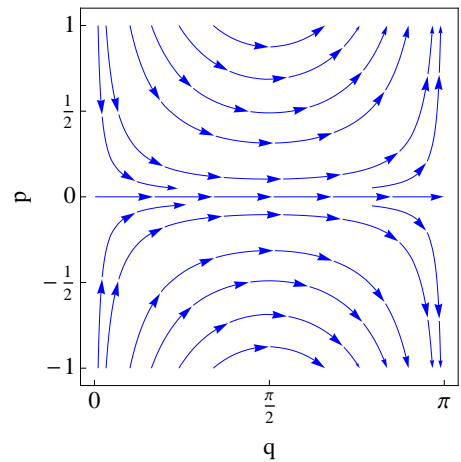}
\caption{Trajectories  in the phase space with original variables
$(q,p)$.} \label{PhasePort1}
\end{figure}

Solutions to these equations are given by
\begin{eqnarray}
\label{phas1} q (T)&=& 2 \arctan \exp (T-T_0),  \\
 \label{phas2} p (T)&=& p_0 \cosh (T-T_0),
\end{eqnarray}
where $p_0$ and $T_0$ are constants of integration. The solution
for $p$ presents a nonsingular symmetric bounce type evolution.
Fig. \ref{PhasePort1} presents the phase portrait of the phase
space, illustrating Eqs. (\ref{phas1}) and (\ref{phas2}).

\subsection{Canonical transformation}

Let us  define the  map $(q,p) \in (0,\pi)\times \dR \mapsto (Q,P)
\in \dR^2$ as follows:
\begin{eqnarray}
Q &:=& \log \tan \left( \frac{q}{2} \right)  \in \mathbb{R}, \label{vQ}\\
P &:=&  p\ \sin(q) \in \mathbb{R}. \label{vP}
\end{eqnarray}
In contrast to  $q$ and $p$, both new variables $Q$ and
$P$ are defined on the entire real line. The variables
$Q$ and  $P$ fulfill the following Poisson
bracket
\begin{eqnarray}
\left\{ Q, P \right\} = \frac{\partial Q}{\partial q}
\frac{\partial P}{\partial p} -\frac{\partial Q}{\partial p}
\frac{\partial P}{\partial q}  = 1 ,
\end{eqnarray}
so the transformation  (\ref{vQ})-(\ref{vP})  is canonical.

In what follows we use the following identities
\begin{equation}\label{dQdq}
\sin q  = \frac{1}{\cosh Q},~~~~\frac{dQ}{dq} = \frac{1}{\sin q} =
\cosh Q ,
\end{equation}
which are useful in further considerations.

\begin{figure}[ht!]
\centering
\includegraphics[width=6cm,angle=0]{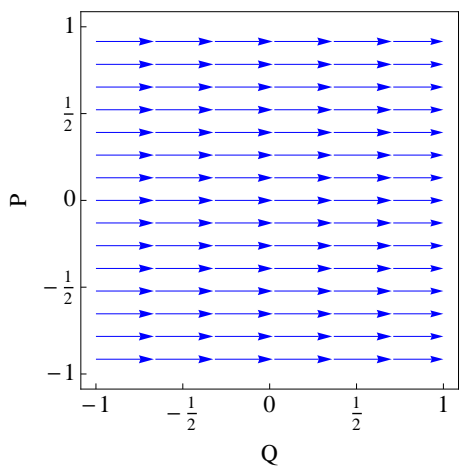}
\caption{Trajectories in the phase space $(Q,P)$.}
\label{PhasePort2}
\end{figure}

In the new variables the Hamiltonian (\ref{Ham}) reads
\begin{equation}
H = P,
\end{equation}
which resembles the Hamiltonian of photon. The equations of motion are:
\begin{eqnarray}
\frac{dQ}{dT} &=& \frac{\partial H}{\partial P} = 1,  \\
\frac{dP}{dT} &=& -\frac{\partial H}{\partial Q} = 0 ,
\end{eqnarray}
which have the solutions
\begin{equation}
Q (T) = T+ c_1,~~~~ P (T) = c_2,
\label{QPsol}
\end{equation}
where $c_1$ and $c_2$ are constants of integration.
The corresponding phase portrait composes of the parallel
trajectories, as shown in Fig. \ref{PhasePort2}.
Comparing Figs. \ref{PhasePort1} and \ref{PhasePort2} we can see
the importance of making suitable choice of phase space variables.
The simplification in describing the dynamics is manifest.

\section{Quantum dynamics}

In order to quantize the system, we introduce the Hilbert space
$\mathcal{H}_1 = L^2([0,\pi],dq)$. In the Dirac LQC, one considers
the Bohr Hilbert space of almost periodic functions, instead of
the $L^2$ space, which complicates quantization.

The action of the operators $\hat{q}$ and $\hat{p}$ is defined as
follows
\begin{eqnarray}
\hat{q} \phi &=& q \phi, \\
\hat{p} \phi &=& -i\hslash \frac{d}{dq} \phi.
\end{eqnarray}
The presence of the symbol $\hslash$ is justified by the physical
dimension of variable $p$ which is an action in the units we have
chosen from the beginning (time has no dimension, and $p$ is
energy-like). While passing from  classical to quantum theory,
along canonical quantization procedure, the known problem of
factor ordering appears. In our case we have
\begin{equation}\label{nfac}
H =  p \; \sin q = \left( \sin q \right)^n p\;\left( \sin q
\right)^{1-n} =: H_n ,
\end{equation}
which can be  quantized as follows (letting aside the formal appearance
of singularities)
\begin{equation}
\hat{H}_n := \frac{1}{2} \left[ \left( \sin\hat{q} \right)^n
\widehat{p}\left( \sin\hat{q} \right)^{1-n} +\left( \sin\hat{q}
\right)^{1-n}  \widehat{p}\left( \sin\hat{q} \right)^{n} \right] ,
\end{equation}
where $n \in \mathbb{R}$.

For such a class of possible factor orderings, the resulting
Hamiltonian takes the following form
\begin{equation}
\hat{H}_n\phi = -i \hslash \left[ \frac{1}{2} \cos q  +\sin q
\frac{d}{dq}
 \right]\phi =: \hat{H}\phi ,
\end{equation}
so in a sense it does not depend on $n$. In some cases, however,
it is advantageous to   make some specific choice. The case $n=1$
was used in Ref. \cite{Mielczarek:2011mx}, while $n=1/2$ is
considered in what follows.  The latter choice is useful while
defining an isometry transformation.

\subsection{Eigenfunctions}

The eigenequation of the Hamilton operator is the following
\begin{equation}
-i \hslash \left[ \frac{1}{2} \cos(q) \phi +\sin (q) \frac{d\phi}{dq}  \right]  = E\phi,
\end{equation}
where $E \in \dR$  belongs to
the continuous spectrum. Solution to this equation is given by
\begin{equation}
\phi_E(q) = \frac{1}{\sqrt{2\pi \hslash}} \frac{\exp \left\{\frac{i}{\hslash} E
\ln \tan \left(\frac{q}{2}\right) \right\}}{\sqrt{ \sin q}}.
\end{equation}
One can verify that
\begin{eqnarray}
\langle \phi_{E_1} |  \phi_{E_2} \rangle &=& \int_0^{\pi}
\overline{\phi_{E_1}(q)} \phi_{E_2}(q) dq  \nonumber  \\
&=& \frac{1}{2\pi \hslash} \int_0^{\pi}   dq \frac{\exp
\left\{\frac{i}{\hslash} (E_2-E_1) \ln \tan \left(\frac{q}{2}
\right) \right\}}{ \sin q}\nonumber  \\
&=& \frac{1}{2\pi \hslash} \int_{-\infty}^{+\infty} dQ
e^{\frac{i}{\hslash}(E_2-E_1)Q } \nonumber \\
&=& \delta(E_2-E_1),
\end{eqnarray}
where we have used the mapping (\ref{vQ})
\begin{equation}
(0,\pi) \ni q \rightarrow Q(q) \in \dR.
\end{equation}

\subsection{Unitary map}

For $n=1/2$, the quantum Hamiltonian reads
\begin{equation}
\hat{H} := \sqrt{\sin\hat{q}}\ \widehat{p}\ \sqrt{\sin\hat{q}}.
\end{equation}
Now, we introduce a new Hilbert space $\mathcal{H}_2 = L^2(
\mathbb{R}, dQ)$ connected with $\mathcal{H}_1$ as follows
\begin{equation}
\mathcal{U}: \  \phi(q) \in  \mathcal{H}_1 \rightarrow \psi(Q) =
\sqrt{\sin(q(Q))} \phi(q(Q)) \in \mathcal{H}_2.
\end{equation}
In particular, the eigenfunctions $\phi_E(q)$ are mapped into the
plane waves
\begin{equation}
\mathcal{U}: \  \phi_E(q) \rightarrow \psi_E(Q) = \frac{e^{\frac{i}{\hslash}
E Q}}{\sqrt{2\pi \hslash}}.
\end{equation}

The map $\mathcal{U}$ is an invertible isometry (unitary
transform), since we have
\begin{eqnarray}
\langle \phi_1 |  \phi_2 \rangle_{\mathcal{H}_1} &=&  \int_0^{\pi}
\overline{\phi_1(q)} \phi_2(q) dq  \nonumber \\
&=& \int_{-\infty}^{+\infty} \overline{\phi_1(q(Q))} \phi_2(q(Q))  \sin(q(Q)) dQ \nonumber \\
&=& \int_{-\infty}^{+\infty}   \overline{\psi_1(Q)} \psi_2(Q) dQ = \langle
\psi_1 |  \psi_2 \rangle_{\mathcal{H}_2}.
\end{eqnarray}
Under this isometry, an operator $\hat{O}_{\mathcal{H}_1} $ acting
in $\mathcal{H}_1$ is transformed into  an operator
$\hat{O}_{\mathcal{H}_2}$ acting on $\mathcal{H}_2$ and vice versa
\begin{eqnarray}
\hat{O}_{\mathcal{H}_2} &=& \mathcal{U} \hat{O}_{\mathcal{H}_1}  \mathcal{U}^{-1}, \label{trans1} \\
\hat{O}_{\mathcal{H}_1} &=& \mathcal{U}^{-1} \hat{O}_{\mathcal{H}_2} \mathcal{U}. \label{trans2}
\end{eqnarray}

As an example, let us consider the operator $\hat{P}$ which acts
in $\mathcal{H}_2$ as follows
\begin{equation}
\hat{P} \psi(Q) = -i \hslash \frac{d}{dQ} \psi(Q).
\end{equation}
Using the transformation  (\ref{trans2}) as well as relation
(\ref{dQdq}), we obtain
\begin{eqnarray}
\hat{P} \psi(Q) &=&  \mathcal{U}^{-1} \hat{P}  \mathcal{U} \phi(q)   =
\frac{1}{\sqrt{\sin q}}\left( -i\hslash \frac{d}{dQ} \right) \sqrt{\sin q} \phi(q)  \nonumber \\
&=&  \frac{1}{\sqrt{\sin q}} \left( -i\hslash \sin q \frac{d}{dq} \right) \sqrt{\sin q} \phi(q) \nonumber  \\
&=& \sqrt{\sin q} \left( -i\hslash \frac{d}{dq} \right) \sqrt{\sin q} \phi(q)  \nonumber \\
&=& \widehat{\sqrt{\sin q}}\ \widehat{p}\ \widehat{\sqrt{\sin q}} \phi(q) \nonumber  \\
&=& \hat{H} \phi(q).
\end{eqnarray}
Thus, the isometry $\mathcal{U}$ transforms the {\it Hamiltonian}
$\hat{H}$ acting in $\mathcal{H}_1$ into a well known  {\it
momentum} operator $\hat{P}$ acting in  $\mathcal{H}_2$.

Using the above results, we define a unitary evolution operator
$\hat{U}$ as follows:
\begin{eqnarray}\label{unit}
\hat{U} \phi(q) := e^{-\frac{i}{\hslash} \hat{H} T}\phi(q) =
e^{-\frac{i}{\hslash} \hat{P} T}\psi(Q) \nonumber \\
= \exp\left\{-T\frac{d}{dQ} \right\}\psi(Q) = \psi(Q-T).
\end{eqnarray}
Therefore, a time evolution in $\mathcal{H}_1$ corresponds to
translation operator in $\mathcal{H}_2$. So the shape of the
probability distribution is preserved in time.

The classical dynamics of the $Q$ variable is $Q=T+c_1$ (See Eq. \ref{QPsol}).
Thus, if the probability distribution is peaked on the classical trajectory at
some given moment in time, it  will trace the classical trajectory
during the whole evolution \cite{Zipfel}.

We should notice at this point that combining translation (\ref{unit}) with phase
modulation, $\Psi(Q) \mapsto e^{\frac{i}{\hslash} P_0 Q}\Psi(Q):=
\Psi_{P_0}(Q)$, leads to the unitary irreducible representation of
the Weyl-Heisenberg group
\begin{equation}
\label{WHgroup}
\Psi(Q) \mapsto e^{\frac{i}{\hslash} P_0 Q}\Psi(Q-T):= \Psi_{P_0,T}(Q)\, ,
\end{equation}
used for constructing coherent states in quantum mechanics (up to
a constant phase factor) \cite{perelomov86,gazeaubook09} and the
Gabor states for time-frequency analysis used in signal processing
\cite{walkerbook08}. The Weyl-Heisenberg action (\ref{WHgroup})
will be at the heart of the construction of the examples presented
in the next section.

\section{Choice of initial state}

In the next three sections we  investigate three representative
initial quantum states:
\begin{itemize}
\item Boxcar state
\item Standard coherent state
\item Soliton-like state
\end{itemize}
Moreover, for each of these states we  investigate Schr\"odinger
cat type superposition in the form
\begin{equation}
\Psi = \frac{N}{\sqrt{2}} \left(\Psi_{P_0}+\Psi_{-P_0}\right)\,
\label{CatState}
\end{equation}
where $P_0$ is the mean value of the $\hat{P}$ operator in the
constituent state $\Psi_{P_0}$ and $N$ is the normalization
factor. It is known that such states are experimentally attainable
(for instance in quantum optics \cite{grangieretal07}). The state
(see Sec. \ref{CatState}) can be viewed also as a superposition of
the two orientations of triads.  Decoherence of two triads
orientations in LQC was recently studied in Ref.
\cite{Kiefer:2012kp}. In what follows we  study such a process in
our framework.

As was already mentioned, quantum dynamics of the considered model
reduces to shifting the initial state in the $Q$ variable:
$\hat{U}(T) \Psi(Q)= \Psi (Q-T)$. Therefore while we have the
initial quantum state $\Psi(Q)$, the corresponding state at the
time $T$ is obtained by replacing $Q \rightarrow Q-T$.  For the
later convenience we define variable $X:= Q-T$, which absorbs all
time dependence of a given quantum state.

To characterize the states under consideration we  study quantum
moments of the operators $\hat{Q}$ and $\hat{P}$ in the given
state. In particular, determination of the mean values $\langle
\hat{Q} \rangle$ and $\langle \hat{P} \rangle$ is crucial to
compare quantum dynamics with the classical phase-space
trajectories. Furthermore, quantum dispersions
\begin{equation}
\sigma_Q := \sqrt{\langle\hat{Q}^2 \rangle-\langle
\hat{Q}\rangle^2 },~~~~\sigma_P := \sqrt{\langle\hat{P}^2
\rangle-\langle \hat{P}\rangle^2},
\end{equation}
as well as covariance
\begin{eqnarray}
C_{QP} := \langle (\hat{Q}-\langle
\hat{Q}\rangle)(\hat{P}-\langle\hat{P}\rangle )\rangle \nonumber \\
=\frac{1}{2}\langle \hat{Q} \hat{P}+\hat{P}\hat{Q} \rangle -
\langle \hat{Q} \rangle\langle \hat{P} \rangle,
\end{eqnarray}
will be a source of information about spreading and squeezing of the quantum states.
Furthermore, by employing dispersions and covariance, one can define the covariance
matrix
\begin{equation}
{\bf \Sigma} :=  \left[\begin{array}{cc}  \sigma_Q^2 & C_{QP} \\
C_{QP}& \sigma_P^2\end{array}  \right].
\end{equation}
Making use of it, one can define  the Schr{\"o}dinger-Robertson uncertainty relation as follows
\begin{equation}
\det {\bf \Sigma} \geq \frac{\hslash^2}{4}.
\end{equation}
The value of  $\det {\bf \Sigma}$ is an important characteristic
of the quantum state, telling us about spread of the state  on the
phase space. One could expect that for small values of $\det {\bf
\Sigma}$, the quantum systems behaves more like a classical one.
However, it is not a general rule. Therefore, other indicators of
semi-classicality should be used.

In the literature, the relative fluctuations
$\sigma_{\mathcal{O}}/\langle \hat{O} \rangle$ are usually
considered as a measure of semi-classicality. Namely, one could
expect that, in the semiclassical limit, quantum fluctuations of
some observable $\hat{O}$ should be much smaller that its mean
value: $\sigma_{\mathcal{O}}/\langle \hat{O} \rangle \ll 1$. Such
a definition has to be however applied with care. Namely, while
denominator $\langle \hat{O} \rangle$ approaches  zero, the
relative fluctuations diverge even, if the quantum dispersion is
very small. But this does not correspond to any strong quantum
effects, and is only a result of the definition of the observable
$\hat{O}$. Another important aspect is the meaning of choosing
$\sigma_{\mathcal{O}}/\langle \hat{O} \rangle$ being \emph{much
smaller} that one. Should it be one hundred or  one million times
smaller? While the choice is quite arbitrary, we already have a
reference value coming from the measurements of a given
observable. If our experimental abilities does not allow us to see
the quantum aspects of a given phenomenon, then we can state that
the underlying dynamics is classical. Therefore, the state can be
called semiclassical if relative fluctuations of some observable
are smaller than the relative uncertainty of experimental
(observational) determination of that observable.

In what follows we  study one quantity, which can be used to
characterize semi-classicality of a quantum state. Namely, the
Wigner function, which is a quasi-probability distribution defined
on the phase space.  Having the wave function $\Psi(Q)=\langle Q |
\Psi \rangle$ of a pure state $| \Psi \rangle$, the Wigner
function is defined to be
\begin{equation}\label{WignerW}
W(Q,P) := \frac{1}{\pi \hslash} \int_{-\infty}^{+\infty}
\overline{\Psi}(Q+y)\Psi(Q-y)e^{2iPy/\hslash} dy.
\end{equation}
The basic properties of the Wigner function are
\begin{eqnarray}
\int_{-\infty}^{+\infty}\int_{-\infty}^{+\infty}W(Q,P)dQ dP = 1, \\
\int_{-\infty}^{+\infty} W(Q,P) dP = |\Psi(Q)|^2, \\
 \int_{-\infty}^{+\infty} W(Q,P) dQ = |\Psi(P)|^2, 
\end{eqnarray}
and
\begin{equation}
-\frac{1}{\pi \hslash} \leq W(Q,P) \leq \frac{1}{\pi \hslash}.
\end{equation}
The last property tells us that also negative values of the the
Wigner function are allowed.  It was suggested that such negative
part of the Wigner function can be considered as an indicator of
quantumness  \cite{Kenfeck2004}. Briefly, less negative the Wigner
function is, more classically the system behaves. In this context
the parameter $\delta(\Psi)$  was introduced in Ref.
\cite{Kenfeck2004}:
\begin{equation}
\delta(\Psi) := \int_{-\infty}^{+\infty} \int_{-\infty}^{+\infty}
|W(X,P)| dX dP -1 \in [0,\infty].
\end{equation}
One half of this parameter is equal to the modulus of the integral
over those domains of the phase space where the Wigner function is
negative.

One says that a state is  semiclassical if $\delta(\Psi) \ll1$.

In this paper, we  study also areas of the negative parts of the
Wigner functions for the considered states. This issue, as far as
we know, was not systematically investigated yet. We show that the
structure of these negative sectors may uncover some deep aspects
of the formulation of quantum mechanics on phase space.

The Wigner function for the Sch{\"o}rdinger cat state (see Sec.
\ref{CatState}) can be written as the following sum
\begin{equation}
W(X,P) = \frac{N^2}{2}(W_{+}+W_{-})+W_{\text{int}},
\end{equation}
where $W_{\pm}$ are Wigner functions for $\Psi_{\pm P_0}$ states while
the interference term
\begin{eqnarray}
W_{\text{int}}&=& \frac{N^2}{2\pi \hslash} \int_{-\infty}^{+\infty}\left[
\overline{\Psi}_{P_0}(x+y)\Psi_{-P_0}(x-y)  \right. \nonumber \\
&+&\left.  \overline{\Psi}_{-P_0}(x+y)\Psi_{P_0}(x-y)  \right]e^{2iPy/\hslash} dy.
\end{eqnarray}

Surprisingly for the considered states, including the
Schr{\"o}dinger cat states, it is possible to find analytical
formulas for the corresponding Wigner functions. Plots of the
Wigner functions will allow to better visualize some quantum
aspects of the states.

\section{Boxcar state}

\subsection{Construction}

We begin with considering the most basic example of a state with
compact support, namely the rectangular or `boxcar' window, widely
used in  Gabor signal analysis \cite{walkerbook08}.  The
definition of the state is the following
\begin{equation}
\Psi_{P_0}(Q) = \left\{  \begin{array}{ccc}  0   & \text{for} &  |Q| >\frac{L}{2} \\
 \frac{1}{\sqrt{L}} e^{\frac{i}{\hslash} Q P_0} & \text{for} &  |Q| \leq \frac{L}{2}
 \end{array}    \right. ,
\end{equation}
where $L>0$ (the dependence of $\Psi_{P_0}$  on parameter $L$ is not made explicit
for the sake of  simplicity). This state can be also written as
\begin{equation}
\Psi_{P_0}(Q) = \frac{1}{\sqrt{L}}  e^{\frac{i}{\hslash} Q P_0} \Theta \left( \frac{L}{2}+Q\right)
\Theta \left( \frac{L}{2}-Q\right),
\end{equation}
where $\Theta(x)$ is the Heaviside step function.

\subsection{Quantum moments}

The mean value and the dispersions of $\hat{Q}$ at the time $T$
are found to be
\begin{eqnarray}
\langle \hat{Q} \rangle  &=& T, \\
\sigma_Q &=& \frac{L}{\sqrt{12}}.
\end{eqnarray}
As expected from the previous analysis, evolution of the mean
value $\langle \hat{Q} \rangle $ traces the classical dynamics.
Here, the quantum dynamics correspond to the  classical trajectory
with the constant of integration $c_1 = 0$ (see  Eq.
(\ref{QPsol}).

Difficulties due to the discontinuous character of the considered
state prevent the evaluation of a similar quantity for the
momentum, since the computation of $\langle \hat{P} \rangle $ and
$\langle \hat{P}^2 \rangle$ involves  divergent integrals. In
principle, one can proceed with a regularization of these
divergences, however the obtained expressions for $\langle \hat{P}
\rangle$ and $\sigma_P$ would not have standard
interpretation. Such results, are however of little use, and
therefore not discussed here.

\subsection{Wigner function}

\begin{figure}[ht!]
\centering
\includegraphics[width=8cm,angle=0]{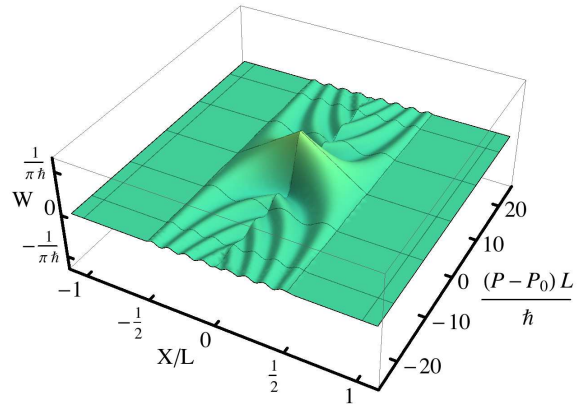}
\caption{Wigner function for  the boxcar state.} \label{WignerA}
\end{figure}

\begin{figure}[ht!]
\centering
\includegraphics[width=6cm,angle=0]{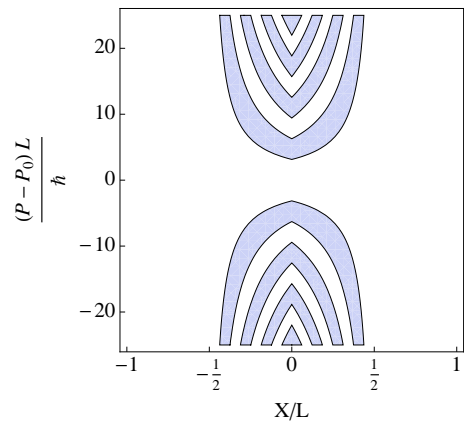}
\caption{Regions where the Wigner function for the boxcar state
assumes its negative values.} \label{Neg1}
\end{figure}

Based on definition (\ref{WignerW}), Wigner function for the boxcar state is
\begin{equation}
W(X,P) = \left\{  \begin{array}{ccc}  0   & \text{for} &  |X| >\frac{L}{2} \\
 \frac{\sin\left[ \dfrac{2(P-P_0)}{\hslash} \left( \frac{L}{2}-|X| \right) \right]}{\pi L (P-P_0)}
 & \text{for} &  |X| \leq \frac{L}{2}    \end{array}    \right.
\end{equation}
We plot this function in Fig. \ref{WignerA}.
Furthermore, in Fig. \ref{Neg1}, we show regions of the negative
values of the Wigner function.

It is worth noticing that for $X=0$ and $P=P_0$ the Wigner function takes the
maximal possible value:
\begin{equation}
W(0,P_0) = \frac{1}{\pi \hslash}.
\end{equation}
As we will see later, analysis of the Wigner functions suggests
that
\begin{equation}
W(\langle \hat{X} \rangle, \langle \hat{P} \rangle) = \frac{1}{\pi \hslash}.
\label{WXP}
\end{equation}
As far as we know such a relation was not proved so far. However,
at least for the known Wigner functions it is always fulfilled.
For the boxcar state, indeed $\langle \hat{X} \rangle = \langle
\hat{Q} \rangle -T= T-T = 0$, however the value of $ \langle
\hat{P} \rangle$ is not properly defined in the present case.
However, since the obtained Wigner function is symmetric with
respect to $P_0$, one can formally write $\langle \hat{P} \rangle
= P_0$, supporting the relation (\ref{WXP}).

\subsection{Schr\"odinger cat state}

The normalization factor for the Schr\"odinger cat  state composed
of two boxcar states is
\begin{equation}
N = \frac{1}{\sqrt{1+\frac{\sin(LP_0/\hslash)}{(LP_0/\hslash)}}}.
\end{equation}

The interference part of the Wigner function is
\begin{equation}
W_{\text{int}} = \left\{  \begin{array}{ccc}  0   & \text{for} &  |X| >\frac{L}{2} \\
\dfrac{N^2 \cos\left[ \frac{2 P_0 X}{\hslash} \right]\sin \left[ \frac{2P}{\hslash}
\left(\frac{L}{2}-|X| \right) \right] }{\pi L P} & \text{for} &  |X| \leq \frac{L}{2}
 \end{array}    \right.
\end{equation}

\begin{figure}[ht!]
\centering
\includegraphics[width=8cm,angle=0]{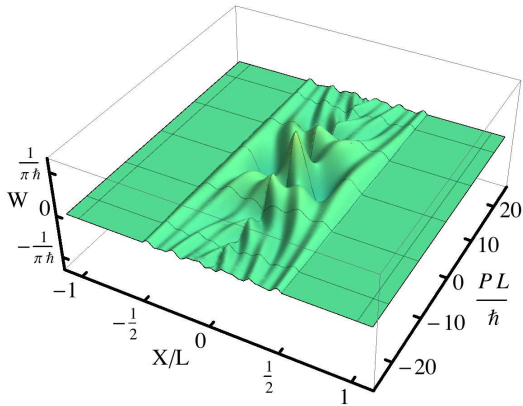}
\caption{Wigner function of the  Schr\"odinger cat state $L P_0 =
7 \hslash$.} \label{WigBoxCat}
\end{figure}

Plot of the Wigner function for the Schr\"odinger cat  state composed
of two boxcar states is shown in Fig. \ref{WigBoxCat}.

\begin{figure}[ht!]
\centering
\includegraphics[width=6cm,angle=0]{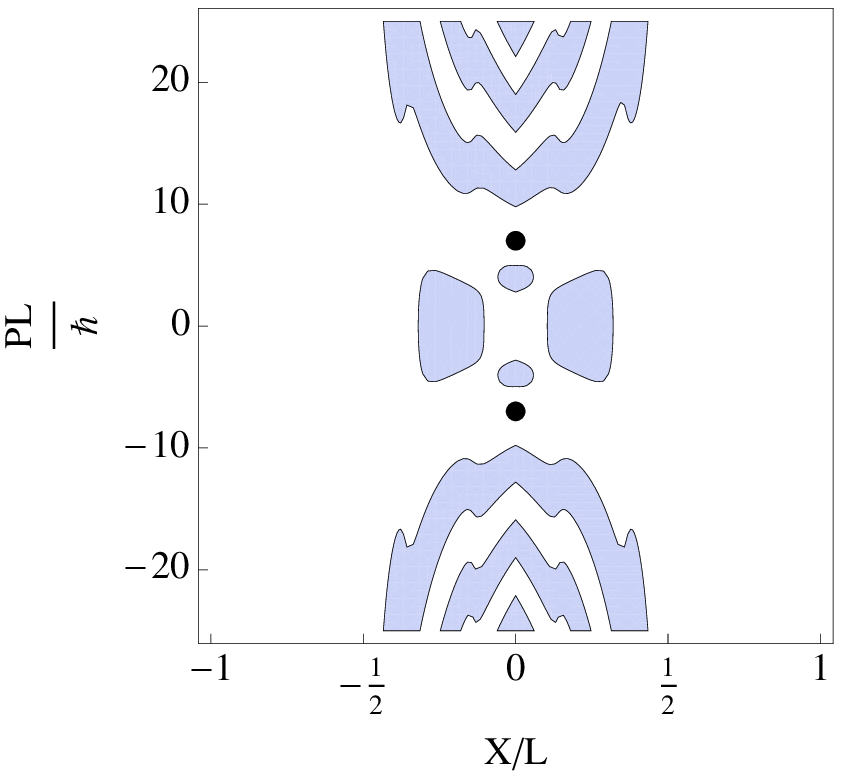}
\caption{Regions where the Wigner function for the Schr\"odinger
cat state assumes its negative values,
 with $L P_0 = 7 \hslash$.  The black dots
represent peaks of the two constituent states.} \label{NegCatBox}
\end{figure}

In Fig. \ref{NegCatBox} we show regions of the negative values of the Wigner function.
For $L P_0 \gtrsim  6.04$ there are two  regions with
negative values of the Wigner function, located at the $X=0$ axis.
We observe that for the particularly case $L P_0 = 7 \hslash$ the
area of each of these regions is about $0.198 \hbar$. However,
while approaching the value $L P_0 \approx 6.04$, the area of
these regions falls to zero. At around $L P_0 \gtrsim 7.64$ these
two regions merge with the two negative domains located at $P=0$
axis.

\section{Standard coherent state}
\label{StandardCoherentState}

\subsection{Construction}

Standard or Glauber coherent states (see \cite{gazeaubook09}
and references therein) are known to play a central role in
studies of connection between classical and quantum world.

In quantum cosmology one expects that in the low energy density
limit, the state is described by the coherent state mimicking the
classical behavior. Here we study the state in which such a
semi-classical behavior is preserved during the whole evolution.
Namely, we consider a squeezed initial wave packet state like
\begin{equation}
\Psi_0(Q) = \int_{-\infty}^{+\infty} f(P) \psi_P(Q) dP ,
\label{Psi0Coh}
\end{equation}
where $\psi_P(Q)$ is the eigenstate of the $\hat{P}$ operator and
\begin{equation}
f(P) =\exp  \left(-z_1P^2+z_2P+z_3\right),
\end{equation}
where $z_1,z_2,z_3 \in \mathbb{C}$, such that $\Re z_1 > 0$. The
integral (\ref{Psi0Coh}) is of the Gaussian type and can be
calculated analytically. Due to (\ref{unit}) we get $\Psi(Q,T) =
\hat{U}\Psi_0(Q)$ which finally reads
\begin{equation}
\Psi(Q,T) = \left( \frac{2 \Re a_1}{\pi} \right)^{1/4}
e^{-a_1(Q-T)^2+a_2 (Q-T)-\frac{(\Re a_2)^2}{4 \Re a_1}}.
\label{cohstate}
\end{equation}
The factors $a_1$ and $a_2$ can be expressed in terms of
coefficients $z_1$ and $z_2$ as follows
\begin{equation}
a_1 := \frac{1}{4\hslash^2 z_1},~~~~~~ a_2 := \frac{i z_2}{2
\hslash z_1}.
\end{equation}
It is worth stressing that the condition $\Re z_1 > 0$ enables
normalization of the state.

\subsection{Quantum moments}

The mean values of the canonical variables in the state
(\ref{cohstate}) are found to be
\begin{equation}
\langle \hat{Q} \rangle = T+\underbrace{\frac{1}{2} \frac{ \Re
a_2}{ \Re a_1}}_{=\ c_1 \in \mathbb{R}},~~~~~~\langle \hat{P}
\rangle = \underbrace{ \hslash  \left( \frac{ \Im a_2}{ \Im a_1}
-\frac{ \Re a_2}{ \Re a_1} \right)\Im a_1 }_ {=\ c_2 \in\mathbb{R}},
\end{equation}
and are in agreement with the classical solutions (\ref{QPsol}).
Furthermore, the dispersions are
\begin{equation}
\sigma_Q  = \frac{1}{2\sqrt{\Re a_1}}
\end{equation}
and
\begin{equation}
\sigma_P  =  \hslash \sqrt{\Re a_1} \sqrt{1+ \left( \frac{ \Im
a_1}{ \Re a_1} \right)^2}.
\end{equation}
Finally,  the covariance reads
\begin{eqnarray}
C_{QP}  = -
\frac{\hslash}{2}  \frac{ \Im a_1}{ \Re a_1} .
\end{eqnarray}
Based on the above, the determinant of the covariance matrix reads

\begin{eqnarray}
\det {\bf \Sigma}  &=& \sigma_Q^2\sigma_P^2 - C_{QP}^2 \nonumber \\
&=& \frac{1}{4\Re a_1}  \hslash^2 \Re a_1
\left[1+ \left( \frac{ \Im a_1}{ \Re a_1} \right)^2\right] - \left(\frac{\hslash}{2}
\frac{ \Im a_1}{ \Re a_1}\right)^2 \nonumber \\
&=& \frac{\hslash^2}{4}.
\end{eqnarray}
As expected for such a case, the  squeezed state saturates
the Schr\"odinger-Robertson uncertainty relation.

\subsection{Wigner function}

For the standard coherent state, the Wigner function takes the
form of the two dimensional Gauss distribution
\begin{equation}
W(X,P) =\frac{1}{\pi \hslash}   \exp \left( -\frac{1}{2} {\bf x}^T
{\bf \Sigma}^{-1} {\bf x} \right),
\label{Wigner}
\end{equation}
where ${\bf x} = (X - \langle \hat{X}  \rangle,  P - \langle
\hat{P}  \rangle )$.  It is worth stressing, that the Wigner
function (\ref{Wigner}) is positive definite.  As Hudson-Piquet
\cite{Hudson1974} theorem says, this is a characteristic property
of the standard coherent states,  distinguishing them among other
pure states. Therefore, positiveness of the Wigner function can be
treated as a definition of the standard coherent state. Taking
negativeness of the Wigner function as a measure of the
quantumness, one can conclude that the standard coherent states
are the most classical pure states. Furthermore, the
positivitiveness of the Wigner function is observed also for the
mixed states \cite{Brocker1995}.

\begin{figure}[ht!]
\centering
\includegraphics[width=8cm,angle=0]{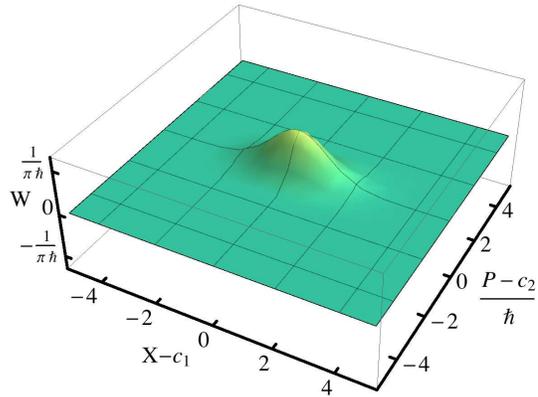}
\caption{Wigner function for the standard coherent state.
We set $C_{QP}=0$ and $\sigma_{Q}=1$, which implies $\sigma_{P}=\hslash/2$.}
\label{WignerB}
\end{figure}

Plot of the Wigner function (\ref{Wigner}) is shown in Fig. \ref{WignerB}.
We set  $C_{QP}=0$, thus axes of symmetry of the obtained Gaussian
distribution overlap with the axes of coordinates. In the case of
$C_{QP}\neq 0$ the distribution would rotate around the centre of
the coordinate system.

\subsection{Schr{\"o}dinger's cat state}

The standard coherent state can be interpreted as the least
quantum state, since they saturate uncertainty relation during the
whole evolution. However, by creating superposition of such
states, strictly non-semiclassical states can be formed. Here, we
study Schr{\"o}dinger's cat type superposition of two standard
coherent states. For simplicity, we consider the case in which
$\Re a_1= \frac{1}{4\hslash^2\alpha}$, $\Im a_1 =0$, $\Re a_2 =0$
and $\Im a_2 = \frac{P_0}{\hslash}$.  Taking this into account,
the state (\ref{cohstate}) reduces to
\begin{equation}
\Psi_{P_0}(X) = \frac{1}{(2\pi \hslash^2 \alpha)^{1/4}} e^{-\frac{X^2}{4\hslash^2
\alpha} +\frac{i}{\hslash} P_0 X}.
\end{equation}
Based on this, we construct Schr{\"o}dinger's cat type superposition \label{CatState}.
The normalization factor in this case is
\begin{equation}
N = \frac{1}{\sqrt{1+\exp(-2P_0^2 \alpha)}},
\end{equation}
thus in the limit $P_0 \rightarrow 0 $ we get $N\rightarrow
1/\sqrt{2}$, so $\Psi \rightarrow \Psi_{P_0=0}$. The corresponding probability distribution
function takes the form:
\begin{equation}
|\Psi(X)|^2 = \frac{N^2e^{-\frac{X^2}{2\hslash^2 \alpha}}}{\sqrt{2\pi \hslash^2 \alpha}}
\left[1+\cos\left(\frac{2 P_0 X}{\hslash}\right)  \right].
\end{equation}

\begin{figure}[ht!]
\centering
\includegraphics[width=7cm,angle=0]{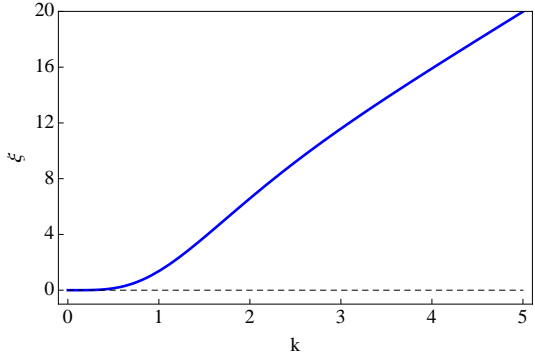}
\caption{Function $\xi(k)$.}
\label{Xik}
\end{figure}

The mean values of the canonical variables in the state (see Sec.
\ref{CatState}) are
\begin{equation}
\langle \hat{Q} \rangle = T,~~~~~~\langle \hat{P} \rangle = 0.
\end{equation}
Therefore, they correspond to the particular classical trajectory with $c_1=0=c_2$.
The dispersions are:
\begin{equation}
\sigma^2_Q = (\hslash^2\alpha) \frac{1+e^{-2k^2}(1-4k^2) }{1+e^{-2k^2}},
\label{sigmaQCat}
\end{equation}
where for the later convenience we have  introduced the
dimensionless parameter $k:=P_0\sqrt{\alpha}$, and
\begin{equation}
\sigma^2_P = \frac{\hslash^2}{4} \frac{1}{(\hslash^2\alpha)} \frac{(e^{-2k^2} +(1+4k^2))}{1+e^{-2k^2}},
\end{equation}
while the covariance is vanishing
\begin{equation}
C_{QP}=0.
\end{equation}
Using the above we have:
\begin{equation}
\sigma^2_Q \sigma^2_P-C_{QP}^2= \frac{\hslash^2}{4}(1+ \xi(k) ) \geq \frac{\hslash^2}{4},
\end{equation}
where the inequality comes from the fact that $\xi(k)\geq 0$. Plot of the function $\xi(k)$ is
shown in Fig. \ref{Xik}.

\begin{figure}[ht!]
\centering
\includegraphics[width=8cm,angle=0]{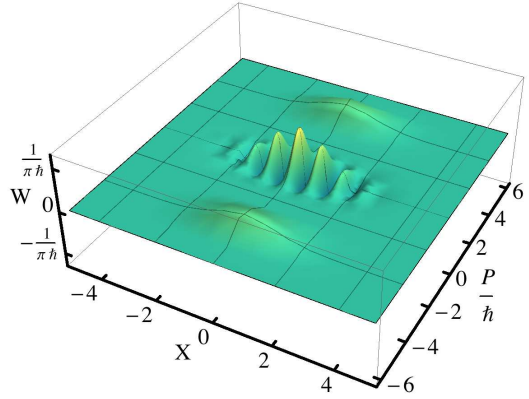}
\caption{Wigner function for the Schr{\"o}dinger cat state.  Here $\alpha = 1
/\hslash^2$ and $P_0 = 4 \hslash$, such that $k=4$.}
\label{Wigner1}
\end{figure}

In the expression for the Wigner function
\begin{equation}
W(X,P) = \frac{N^2}{2}(W_{+}+W_{-})+W_{\text{int}},
\end{equation}
we have
\begin{equation}
W_{\pm} =  \frac{1}{\pi \hslash}e^{-\frac{X^2}{ 2\hslash^2
\alpha}-2(P\mp P_0)^2\alpha},
\end{equation}
and the interference term reads
\begin{equation}
W_{\text{int}} = \frac{N^2}{\pi \hslash} e^{-\frac{X^2}{ 2\hslash^2 \alpha}-2P^2\alpha}
\cos\left( \frac{2P_0 X}{\hslash} \right).
\end{equation}
Collecting the contributions, the Wigner function can be written as
\begin{eqnarray}
W(X,P) = \frac{N^2}{\pi \hslash} e^{-\frac{X^2}{ 2\hslash^2 \alpha}-2P^2\alpha} \times\nonumber  \\
\times\left[\cosh(4PP_0 \alpha)e^{-2P_0^2\alpha}
+\cos\left( \frac{2P_0 X}{\hslash} \right)  \right].
\label{WignerCat}
\end{eqnarray}
The plot of this function is shown in Fig. \ref{Wigner1}.

\begin{figure}[ht!]
\centering
\includegraphics[width=7cm,angle=0]{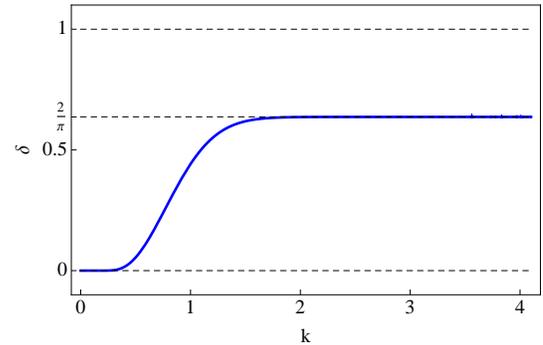}
\caption{Indicator $\delta(\Psi)$ for the Schr{\"o}dinger cat state. In the limit
$k\rightarrow \infty$ the value $\tfrac{2}{\pi} \approx 0.637$ is approached, while
for $k\rightarrow 0$ the state becomes coherent $\delta(\Psi)=0$.}
\label{deltaPsiCat}
\end{figure}

Using the expression (\ref{WignerCat}) we get:
\begin{equation}
\delta(\Psi)  = N^2-1+\frac{N^2}{\sqrt{2\pi}} \int_{-\infty}^{+\infty}
e^{-\frac{z^2}{2}}\left| \cos(2kz) \right| dz.
\end{equation}
We plot this function in Fig. \ref{deltaPsiCat}.
One can show that there exists the following limit:
\begin{equation}
\lim_{k\rightarrow \infty} \delta(\Psi)  = \frac{2}{\pi} \approx 0.637.
\end{equation}

\begin{figure}[ht!]
\centering
\includegraphics[width=6cm,angle=0]{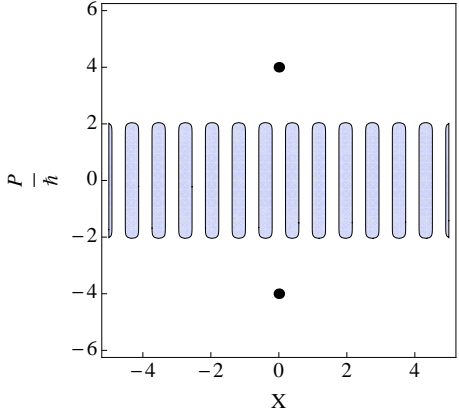}
\caption{Regions where the Wigner function for the CS cat state
assumes its negative values. The black dots represent peaks of the
two constituent states.} \label{SCCSneg}
\end{figure}

The regions of the negative values of the Wigner function are
specified by the inequality:
\begin{equation}\label{pole}
\cosh(4PP_0 \alpha)e^{-2P_0^2\alpha}+\cos\left( \frac{2P_0 X}{\hslash} \right) < 0.
\end{equation}
Due to the periodicity of the cosine function we obtain an
infinite set of ellipse-like regions of negative values of
the Wigner function, as shown in Fig. \ref{SCCSneg}.
The area  of each such region can be expressed as follows
\begin{equation}
\label{area1}
A = \frac{\hslash}{2} \frac{e^{-2k^2}}{k^2} \int_1^{e^{2k^2}}
\frac{\text{arccosh}(z)}{\sqrt{1-e^{-4k^2}z^2}}dz \approx \pi \frac{\hslash}{2} =\frac{h}{4}.
\end{equation}
The value of above integral was found numerically for a broad
range of the parameter $k$, exhibiting independence on the value
of $k$. While we were not able to determine the above integral
analytically, in the $k\rightarrow 0$ limit, after change of
variables, it reduces to
 \begin{equation}
\lim_{k\rightarrow 0} A = \hslash \int_0^1
\frac{d z}{\sqrt{1-z^2}} =  \hslash \frac{\pi}{2}=\frac{h}{4},
\end{equation}
in agreement with the numerical result. Thus, while passing to the
case of the standard coherent states the area of the negative
parts of the Wigner function is preserved. This may seem to be
inconsistent since there are no regions of  negative values for
the coherent states at all. However, this discrepancy is only
apparent. In fact, in the limit $k\rightarrow 0$ the domains of
$W<0$  elongate in the P direction and disperse in the X
direction. In the limit $k\rightarrow 0$, separation between these
domains tends to infinity. In short, the domains of $W<0$ escape
to infinity in the limit $k\rightarrow 0$, maintaining its areas.

It is interesting to notice that the area  (\ref{area1}) is the
same as in the case of the $|1\rangle$ Fock state of the harmonic
oscillator. This coincidence may exhibit some deeper properties of
the formulation of quantum mechanics on  phase space
\cite{schleichbook01,zachosetalbook06}.

\section{Soliton-like state}

\subsection{Construction}

In this section we study a state given by superposition of the eigenstates
of the Hamiltonian with the profile
\begin{equation}
f(P) \propto \frac{1}{ \cosh \left( a(P-P_0) \right)}.
\end{equation}

We will see that while this profile looks qualitatively similar to
the  Gaussian distribution, its properties are significantly
different. In particular, the corresponding Wigner function takes
negative values. We have
\begin{equation}
\Psi_{P_0}(Q) = \sqrt{\frac{\pi}{4 a \hslash}}
\frac{e^{\frac{i}{\hslash} P_0 Q} } {\cosh\left(
\frac{\pi}{2a\hslash} Q\right)} .\label{PsiSoliton}
\end{equation}

This state is also of the hyperbolic secant form, since the
hyperbolic secant function is a fixed point of the Fourier
transform, as the Gaussian distribution.

It is worth mentioning that the obtained state has the form of the
so-called bright soliton, which is a solution of the
Gross-Pitaevskii equation describing Bose-Einstein condensate. For
experimental evidences of such soliton states see for instance
\cite{beckeretal08,Khayk,Cornish}.

The mean values of the canonical variables are
\begin{equation}
\langle \hat{Q} \rangle = T,~~~~~~\langle \hat{P} \rangle = P_0.
\end{equation}
Therefore the classical constant of integration $c_1$ is fixed to be zero,
while integration constant $c_2= P_0$.

The covariance is equal to
\begin{equation}
C_{QP}= 0.
\end{equation}

The dispersions are
\begin{equation}
\sigma_Q =  \frac{a \hslash}{\sqrt{3}},
\end{equation}
and
\begin{equation}
\sigma_P = \frac{1}{a\hslash} \cdot \frac{\pi}{\sqrt{3}} \cdot \frac{\hslash}{2}
\end{equation}

Based on the above, the uncertainty relation is satisfied
\begin{equation}
\sigma_Q\sigma_P = \frac{\pi}{3} \cdot \frac{\hslash}{2} \simeq  1.0472 \cdot
\frac{\hslash}{2} > \frac{\hslash}{2}\,.
\end{equation}
The uncertainty differs only by the factor $\frac{\pi}{3} \approx
1.0472$ from the case of minimal uncertainty realized by the
standard coherent states.

\subsection{Wigner function}

\begin{figure}[ht!]
\centering
\includegraphics[width=8cm,angle=0]{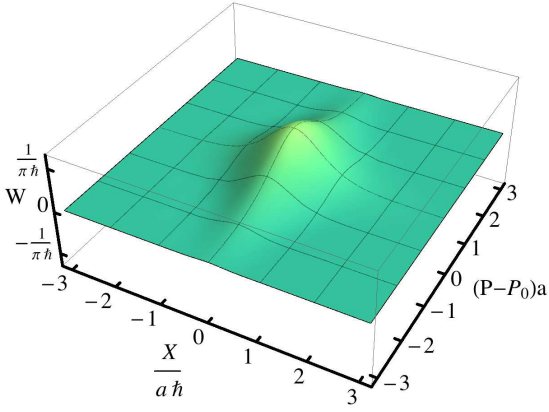}
\caption{Wigner function for  the soliton-like state.}
\label{WignerD}
\end{figure}

Inserting  the wave function (\ref{PsiSoliton}) into the
definition of the Wigner function we  obtain the  integral
\begin{equation}
W(X,P) = \frac{1}{a\hslash^2} \int_0^{\infty} \frac{\cos\left(  \frac{2y}{\hslash}(P-P_0) \right) dy}
{\cosh\left( \frac{\pi}{a\hslash} X\right)+\cosh\left( \frac{\pi}{a\hslash} y\right)}.
\end{equation}
This integral can be calculated analytically by using the formula
(see Integral 3.983 in Ref. \cite{Gradshteyn})
\begin{equation}
\int_0^{\infty} \frac{\cos (a y)}{b \cosh (\beta y)+c} dy =\frac{\pi \sin \left(\frac{a}{\beta}
\text{arccosh}
\left( \frac{c}{b} \right) \right)}
{ \beta \sqrt{c^2-b^2} \sinh\left( \frac{a\pi}{\beta} \right)},
\end{equation}
which holds for $c>b>0$, leading to
\begin{equation}
W(X,P) = \frac{1}{\hslash}   \frac{ \sin \left( \frac{2(P-P_0)}{\hslash} X \right) }{
\sinh (2(P-P_0)a )\sinh\left( \frac{\pi}{a\hslash} X \right)}.
\end{equation}
We plot this function in Fig. \ref{WignerD}.

\begin{figure}[ht!]
\centering
\includegraphics[width=6cm,angle=0]{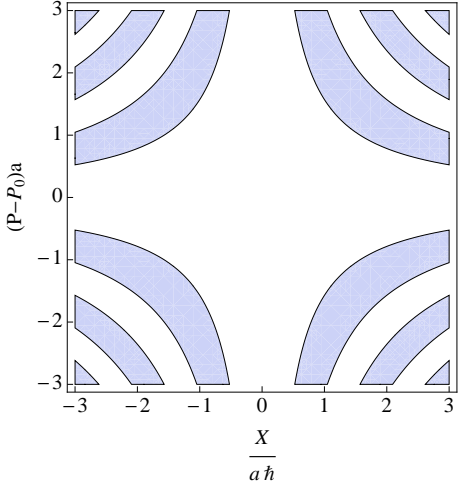}
\caption{Regions where the Wigner function for the soliton-like
state assumes its negative values. }
\end{figure}

By employing the integral
\begin{equation}
\int_0^{\infty} \frac{\sin(\alpha y)}{\sinh (\beta y)}dy = \frac{\pi}{2\beta}
\tanh\left(\frac{\alpha \pi}{2 \beta} \right)
\end{equation}
one can verify that
\begin{equation}
\int_{-\infty}^{+\infty} W(X,P) dP = \frac{\pi}{4 a \hslash}
\frac{1} {\cosh^2\left(  \frac{\pi}{2a\hslash} X\right)}
\end{equation}
as expected.

\subsection{Superposition of soliton-like  states}

\begin{figure}[ht!]
\centering
\includegraphics[width=8cm,angle=0]{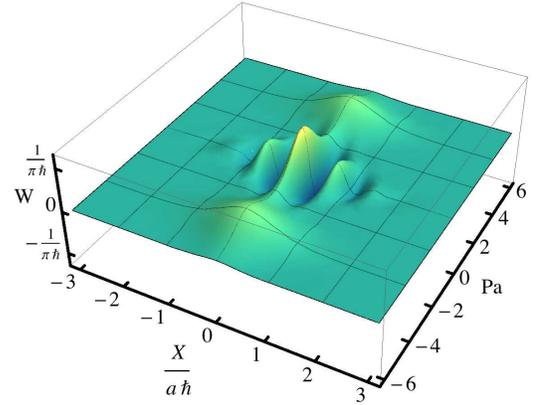}
\caption{Wigner function for the superposition of soliton-like states for $P_0 a=4$.}
\label{WignerE}
\end{figure}

\begin{figure}[ht!]
\centering
\includegraphics[width=6cm,angle=0]{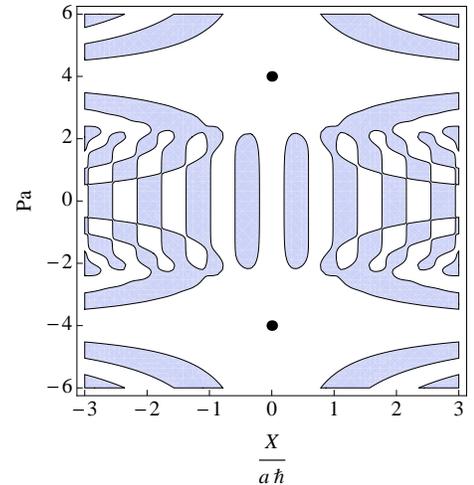}
\caption{Regions where the Wigner function for the soliton-like
cat state assumes its negative values. The black dots represent
peaks of the two constituent states.} \label{NegCSSS}
\end{figure}

The normalization function is equal to
\begin{equation}
N= \frac{1}{\sqrt{1+\frac{2P_0 a}{\sinh(2P_0a)}}}.
\end{equation}
The interference part of the Wigner function is found to be
\begin{equation}
W_{\text{int}} =  \frac{N^2}{\hslash}   \frac{ \cos \left( \frac{2P_0}{\hslash} X \right)
\sin \left( \frac{2P}{\hslash} X \right) }{
\sinh (2Pa )\sinh\left( \frac{\pi}{a\hslash} X \right)}.
\end{equation}
Figure \ref{WignerE} presents the Wigner function.
In Fig. \ref{NegCSSS} we show negative domains of the Wigner function.
The areas of the finite domains of $W<0$ are $A \approx 1.00288 \cdot \tfrac{h}{4} > \tfrac{h}{4}$.

\section{Decoherence}

In quantum mechanics, we are dealing with quantum systems,
observers and environments. Because there is always some
interaction between quantum system and its environment, quantum
systems are never perfectly isolated. Due to this interaction,
quantum systems form an entangled state with its environment. The
environmental degrees of freedom are however inaccessible to
observer. Therefore, from the observer point of view, a quantum
system is described by a mixed quantum state, being a statistical
mixture of so-called \emph{pointer states}. These
\emph{einselected} states have properties closest to the classical
realm. This process of interaction  of the quantum system with its
environment, leading to emergence of the classical behavior, is
called \emph{decoherence}
\cite{Zurek:1991vd,Zurek:2003zz,Zurek:1992mv}.

In this paper we consider a minisuperspace cosmological model with
two-dimensional phase space, parametrized by $Q$ and $P$. It
describes  dynamics of a global degree of freedom of the universe
- the scale factor.  At the quantum level, state of this system
was described by a pure state in the Hilbert space
$\mathcal{H}_S=L^2(\mathbb{R},dQ)$.

Phase space of the Universe is however much richer and forms the
so-called superspace. There is an infinite number of degrees of
freedom describing the inhomogeneities of the gravitational and
matter fields. These degrees of freedom (irrelevant degrees of
freedom) can serve as an environment for our system (relevant
degrees of freedom). It was tacitly assumed here that the
inhomogeneities have no influence on the background dynamics.
However, this assumption can be violated at some stages of the
cosmic evolution.

The state of our system and its environment is described by a
vector in the Hilbert space which is the tensor product
$\mathcal{H}_S\otimes \mathcal{H}_E$,  where $ \mathcal{H}_E$ is
the Hilbert space of the environment states. It is worth noticing
here that the total state describing the system plus environment
is pure, since there is no environment with respect to the
Universe.  However, from the perspective of an internal observer,
the state of the quantum system, immersed in the inaccessible
environment, is mixed.

Let us consider a possible situation, which can arise in the
cosmological context. Namely let us assume that the state of our
minisuperspace model is described by the Schr\"odinger cat state
composed of two coherent states, as was studied in Sec.
\ref{StandardCoherentState}. Due to interaction with environment,
this state forms an entangled state with its environment. This
state is described by the total density matrix $\hat{\rho}_{E+S}$.
However, for an observer, only elements of the reduced density
matrix $\hat{\rho}_{S} = \text{tr}_E(\hat{\rho}_{E+S})$ are
available, which forms  a statistical mixture of two coherent
states \begin{equation} \hat{\rho}_E = \frac{1}{2} | P_0 \rangle
\langle P_0|+\frac{1}{2} | -P_0 \rangle \langle -P_0|.
\label{reducedmatrix}
\end{equation}
Therefore, while  the initial pure state contained interference
terms in its density matrix \begin{equation} \hat{\rho} = | \Psi
\rangle \langle \Psi |,
\end{equation}
they were suppressed by interaction with the environment.

The described process of decoherence can be clearly seen at the level of the Wigner functions.
Namely, in this process the interference part of the Wigner function is suppressed leading
to the statistical mixture of two coherent states.

\begin{figure}[ht!]
\centering
\includegraphics[width=8cm,angle=0]{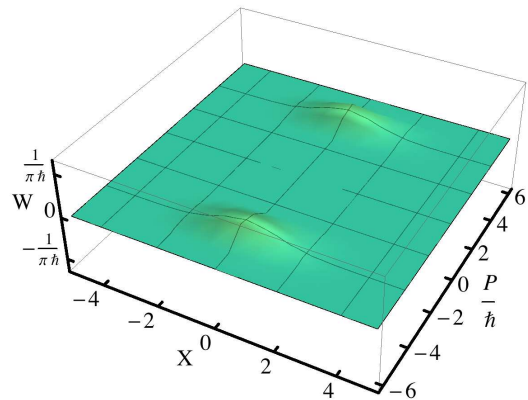}
\caption{Wigner function for the decohered Schr{\"o}dinger cat state.}
\label{WignerCat2}
\end{figure}

Employing the general definition of the Wigner function, which
applies also to  mixed states:
\begin{equation}
W(X,P) = \frac{1}{\pi \hslash} \int_{-\infty}^{+\infty} \langle
X-y | \hat{\rho} | X+y \rangle e^{2iPy/\hslash} dy ,
\end{equation}
we get
\begin{equation}
W(X,P) =\frac{1}{2}(W_{+} +W_{-}),
\end{equation}
for the reduced density matrix (\ref{reducedmatrix}). We show this function in Fig. \ref{WignerCat2}.
By comparing Fig. \ref{WignerCat2} with  Fig. \ref{Wigner1}, it is clear that during
the decoherence process the interference pattern disappears, and the state of
the universe reduces to the statistical mixture of the two uncorrelated universes.

We suggest two possible interpretations of the observed emergence of two separated universes:
\begin{itemize}
\item The two signs of the $P$ variable may be related with the
two possible orientations of triad. Therefore, the Schr\"odinger
cat state may describe universe being in superposition of two
orientations of the triad. These orientations correspond to the
two values  $\pm P_0$. Due to the interaction with environment,
the state of the universe breaks into a statistical mixture of the
two universes with positive and negative orientations,
respectively. Such possibility was recently studied in Ref.
\cite{Kiefer:2012kp}.  It was shown that fermionic matter serves
as a natural environment with respect to which decoherence of the
triad orientation may occur.

\item According to the BKL scenario \cite{BKL22,BKL33}, the
space-points decouple while approaching the cosmic singularity.
Such a behavior, known also as the asymptotic silence, appears
already at the classical level, but is also predicted within some
approaches to quantum gravity. In particular, recent
investigations of the perturbative sector of LQC exhibited
occurrence of the asymptotic silence due to the discrete nature of
space at the Planck scale \cite{Mielczarek:2012tn}.

In the state of asymptotic silence, each small neighbourhood of a point of space can be described
by a homogenous model. In particular, in the BKL scenario, each such small homogeneous
`universe' is described by the Kasner solution. Turning on some nonvanishing coupling between the
different space points leads to transitions between Kasner phases for each of those small
universes.

In the idealized situation, each of the space points may be
described by the homogeneous and isotropic model as the one
considered in this paper. The Schr\"odinger cat state describes
the simplest superposition of two such small universes with
opposite values of $P_0$  respectively.

\end{itemize}

\section{Entropies}

Due to the possible relation  to the flow of time, it is tempting
to introduce the notion of entropy in quantum cosmology. In
particular, in our earlier paper \cite{Mielczarek:2012qs}, the
following definition of entropy of squeezing was proposed:
\begin{equation}
S := k_{B} \ln \left( \frac{\sigma_Q \sigma_P\sqrt{1-\rho^2}}{\hslash/2}\right),
\label{entropy1}
\end{equation}
where $\rho=C_{QP}/(\sigma_Q \sigma_P) \in[-1,1] $ is  a
dimensionless correlation coefficient, measuring phase squeezing
of a quantum state. This entropy is defined such that it is equal
to zero for the states saturating the Schr\"odinger-Robertosn
uncertainty relation and is positive definite for other pure
states.

It was shown that while considering dynamics of  a standard
coherent state (Gaussian state) in the $\mathcal{H}_1$ Hilbert
space, qualitative behavior of (\ref{entropy1}) agrees with
predictions based on the von Neumann entropy of mixed states:
\begin{equation}
S_{N} = -k_B \text{tr} [ \hat{\rho} \ln \hat{\rho} ].
\end{equation}
In particular,  the entropy (\ref{entropy1}) assumes its minimal
value at the  bounce, when  the energy density reaches its maximal
value  \cite{Mielczarek:2012qs}.

The density matrix is invariant under unitary transformation,
$\hat{\rho} \rightarrow \hat{U} \hat{\rho} \hat{U}^{\dagger}$,
which  amounts to unitary invariance of the von Neumann entropy.
This reflects the fact that, at the classical level, Gibbs entropy
as well as Boltzmann entropy  are invariant under canonical
transformations.

The entropy (\ref{entropy1}) is however not invariant with respect
to the canonical transformations. For the variables $(q,p)$  and
$(Q,P)$ related by the canonical transformation, the entropy can
take a completely different form. In particular, while in  one
coordinate system it is time dependent function, in another one it
takes a constant value in time.  Such  a situation happens e.g. in
the case of standard  coherent states considered in this paper. In
Ref. \cite{Mielczarek:2012qs}  we have studied  the evolution of
dispersions $\sigma_q$, $\sigma_p$ and $C_{qp}$ for such a state.
All of these variables exhibited time dependence and both
amplitude and phase squeezing in time  were observed.

However, as shown in Sec. \ref{StandardCoherentState} the values
of $\sigma_Q$, $\sigma_P$ and $C_{QP}$ are constant in time.
Moreover, also for the other states in this paper,   time
independence of dispersions and covariance was observed. This
turns out to be a general property, resulting from the particular
form of the Hamiltonian $H=P$. For such  a system, the time
evolution manifests as a shift of  the Q variable, $Q\rightarrow
Q-T$. Therefore, the whole dynamics reduces to translations of the
Wigner function with preservation of its shape.  Such translation
does not affect the initial spreads and covariance, as well as the
higher moments.  This behavior, is precisely the same as observed
for the photon wave-packets, for which dispersions are preserved
in time due to linear dispersion relation.

Due to the above property, the  Wehrl entropy \cite{wehrl}, which
employs the Husimi \cite{husimi} function, is also expected to
take constant values for the state considered in this paper.
However, it is worth stressing that the von Neumann's entropy
grows during decoherence of the Schr\"odinger cat state studied in
the previous section. This increase of the  von Neumann entropy
exhibits the fact that the part of information about the system is
hidden in the environment.  Such an increase of entropy can
contribute to the total increase of the entropy of the universe
and the observed arrow of time.

\section{Relation to observations}

Each of the states considered in this paper aims to describe
global degrees of freedom of the universe in the quantum epoch.
There is no theoretical principle which could be used to
distinguish one of these states in the context of quantum
cosmology. Such a goal can be achieved only by confronting
theoretical predictions with observational data. The task is
however extremely difficult due to the lack of well established
probes of the Planck epoch. Nevertheless, a significant progress
in this direction has been made in  recent years. Especially, by
employing scalar and tensor perturbations which can be produced
during the quantum epoch. They may lead to some imprints on the
spectrum of the cosmic microwave background (CMB) radiation. In
particular, by using observations of the CMB one can determine the
value of the Hubble factor
\begin{equation}
\mathcal{H} :=\frac{1}{a} \frac{da}{dt},
\end{equation}
during the phase of inflation, which follows the quantum epoch.
This measurement can be used to put constraints on the quantum
fluctuations of the Hubble factor in a given state describing the
Universe. Namely, we expect that \emph{relative quantum
fluctuations of the Hubble parameter are smaller than the relative
uncertainty of the measurement}. It means that:
\begin{equation}
\frac{\sigma_{\mathcal{H}}}{\langle \hat{\mathcal{H}} \rangle} <
\frac{\Delta \mathcal{H}}{\mathcal{H}},
\label{ConstHubb}
\end{equation}
where $\mathcal{H}$ is the measured value of the Hubble parameter
and $\Delta \mathcal{H}$ is the uncertainty of the measurement. As
shown in Ref. \cite{Mielczarek:2012qs}, based on the seven years
observations of the WMAP satellite, uncertainty of the measurement
of the Hubble at some fixed point of inflation is  $\frac{\Delta
\mathcal{H}}{\mathcal{H}}\approx 0.19$. Whereas, measurements of
the present value of the Hubble factor give us $\frac{\Delta
\mathcal{H}}{\mathcal{H}}\approx 0.02$. Both results are in
agreement with our expectation that relative quantum fluctuations
should satisfy
\begin{equation}
\frac{\sigma_{\mathcal{H}}}{\langle \hat{\mathcal{H}} \rangle} \ll 1.
\label{semi1}
\end{equation}
Such a restriction is usually considered as a condition of
semi-classicality \cite{Corichi:2007am}.

One can also interpret the above as a requirement of the
correspondence principle. Namely, in the limit of large quantum
numbers the quantum mechanics should reproduce classical
dynamics. Here, the limit of the large quantum numbers corresponds
to the limit of large volume.

Based on the above we can guess, as confirmed by the astronomical
observations, that the relative fluctuations of the Hubble
parameter may be used to indicate semi-classicality of the
expanding universe. In what follows we  use this restriction to
put constraints on the parameters of considered states of the
universe.

The important observation is that in the limit of large volumes
$(|p| \rightarrow \infty, q\rightarrow 0)$, the value of the
parameter $q$ is proportional to the Hubble factor: $q = \gamma
\lambda \mathcal{H}$. But this is precisely where our
observational constraints can be applied.  Therefore,  in the
considered limit, the constraint (\ref{ConstHubb}) leads to
\begin{equation}
\frac{\sigma_{q}}{\langle \hat{q} \rangle} < \frac{\Delta \mathcal{H}}{\mathcal{H}}.
\end{equation}
The task is now to determine the left hand side of the above
equation for the examined states in the limit $q \rightarrow 0$.
Since we  investigated properties of the states in the Hilbert
space $\mathcal{H}_2$, we have to express the parameter $q$ in
terms of $Q$. Employing the definition (\ref{vQ}), we can write
\begin{equation}
q = 2 \text{arctan}\left( e^{X+T}\right) =\epsilon e^{X} +\mathcal{O}(\epsilon^3),
\end{equation}
where we have performed expansion in the parameter $\epsilon = e^{T}$, which tends to zero in the
limit $T\rightarrow - \infty$. In this limit $q\rightarrow 0$.

Now we express the first and the second moment of the variable $q$ as follows:
\begin{eqnarray}
\langle \hat{q} \rangle &=& 2\epsilon \int_{-\infty}^{+\infty} e^{X}
|\Psi(X)|^2 dX +\mathcal{O}(\epsilon^3), \\
\langle \hat{q}^2 \rangle &=& 4 \epsilon^2 \int_{-\infty}^{+\infty} e^{2X}
|\Psi(X)|^2 dX +\mathcal{O}(\epsilon^4).
\end{eqnarray}
Using  the above expressions, the relative uncertainty of $q$ in
the limit $\epsilon \rightarrow 0$ can be written as
\begin{eqnarray}
\lim_{\epsilon \rightarrow 0} \frac{\sigma_{q}}{\langle \hat{q} \rangle} &=&
\lim_{\epsilon \rightarrow 0} \sqrt{ \frac{\langle \hat{q}^2 \rangle }
{\langle \hat{q}\rangle^2}-1} \nonumber \\
&=& \sqrt{\frac{\int_{-\infty}^{+\infty} e^{2X} |\Psi(X)|^2
dX}{\left(\int_{-\infty}^{+\infty} e^{X} |\Psi(X)|^2
dX\right)^2}-1}.
\label{LimRelFluc}
\end{eqnarray}
The condition of semi-classicality (\ref{semi1}) now states that:
\begin{equation}
\lim_{\epsilon \rightarrow 0} \frac{\sigma_{q}}{\langle \hat{q}\rangle} \ll 1 .
\label{semi2}
\end{equation}
In what follows, we will use this restriction to put constraints on the parameters of our models.

\subsection{Boxcar state}

For the boxcar state, the following integrals can be easily found:
\begin{eqnarray}
\int_{-\infty}^{+\infty} e^{X} |\Psi(X)|^2 dX &=& \frac{2}{L} \sinh \left( \frac{L}{2} \right),\\
\int_{-\infty}^{+\infty} e^{2X} |\Psi(X)|^2 dX &=&  \frac{1}{L} \sinh (L).
\end{eqnarray}
By applying those integrals to expression (\ref{LimRelFluc}), we obtain
\begin{equation}
\lim_{\epsilon \rightarrow 0} \frac{\sigma_{q}}{\langle \hat{q} \rangle} = \sqrt{ \frac{L/2}{\tanh(L/2)}-1}.
\label{RelFlucLimBS}
\end{equation}
Employing the condition of semi-classicality (\ref{semi2}) we obtain the following
transcendental ineqality
\begin{equation}
2\tanh(L/2) \gg L/2.
\end{equation}
This equation can be solved numerically leading to the restriction
\begin{equation}
L \ll 3.83.
\label{L383}
\end{equation}
Only for such values the state reveals the correct semi-classical
limit.  As we see, by imposing the requirement of an agreement
with the semi-classical behavior of the universe in the expanding
branch, a dominant part of the range of the parameter $L$ is
excluded.

\subsection{Schr\"odinger cat composed of boxcar states}

\begin{figure}[ht!]
\centering
\includegraphics[width=7cm,angle=0]{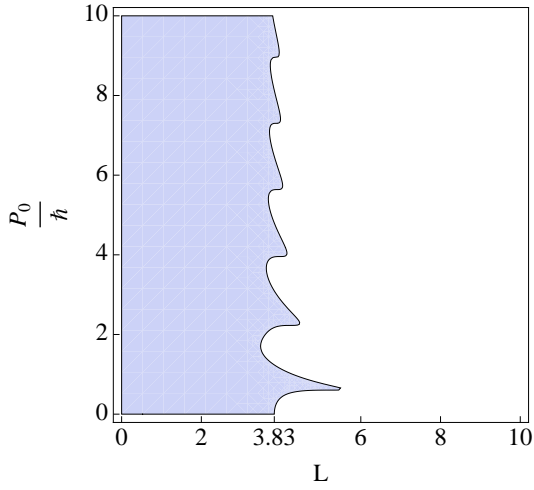}
\caption{Shadowed region of the parameter space represents
$\lim_{\epsilon \rightarrow 0} \frac{\sigma_{q}}{\langle
\hat{q}\rangle} < 1$, a necessary condition for semi-classicality,
which holds in a neighborhood of the origin.} \label{SemiRegion1}
\end{figure}

For the Schr\"odinger cat composed of boxcar states, the formula (\ref{LimRelFluc})
can be expressed as follows
\begin{widetext}
\begin{equation}
\lim_{\epsilon \rightarrow 0} \frac{\sigma_{q}}{\langle \hat{q} \rangle} = \sqrt{  \frac{
 e^L (1 + 4 \tilde{P}_0^2)^2 (L \tilde{P}_0 + \sin(L \tilde{P}_0)) (\tilde{P}_0 \cosh(L) \sin(L \tilde{P}_0)
 +(1 + \tilde{P}_0^2 + \cos(L \tilde{P}_0)) \sinh(L))}{((\tilde{P}_0 + \tilde{P}_0^3) ((-1 + e^L) (1 + 4 \tilde{P}_0^2)
 + (-1 + e^L) \cos(L \tilde{P}_0) + 2 (1 + e^L) \tilde{P}_0 \sin(L \tilde{P}_0))^2)} -1},
\label{RelFlucLimSBS}
\end{equation}
\end{widetext}
where $\tilde{P_0}=P_0/\hslash$. This equation simplifies to
(\ref{RelFlucLimBS}), in the limit $\tilde{P_0}\rightarrow 0$.
Based on equation (\ref{RelFlucLimSBS}), in Fig. \ref{SemiRegion1}
we show the region of the parameter space for which
$\lim_{\epsilon \rightarrow 0} \frac{\sigma_{q}}{\langle
\hat{q}\rangle} <1$. In this region, relative quantum fluctuations
of the Hubble factor in the large volume limit are smaller than
unity.
Therefore, only for the values of the parameters belonging to this
region, the Schr\"odinger cat composed of boxcar states has a
semiclassical limit. It is worth noticing that for $P_0/\hslash =
0$, the obtained constraints overlap with (\ref{L383}), as
expected.

\subsection{Standard coherent state}

For the standard coherent state
\begin{eqnarray}
\int_{-\infty}^{+\infty} e^{X} |\Psi(X)|^2 dX &=&  \exp \left[  \frac{ (2\Re a_2+1)^2}{8 \Re a_1} \right],\\
\int_{-\infty}^{+\infty} e^{2X} |\Psi(X)|^2 dX &=&  \exp \left[  \frac{ (2\Re a_2+2)^2}{8 \Re a_1} \right].
\end{eqnarray}
Using these integrals we get
\begin{equation}
\lim_{\epsilon \rightarrow 0} \frac{\sigma_{q}}{\langle \hat{q} \rangle} =
\sqrt{ \exp\left[\frac{1-2(\Re a_2)^2}{4\Re a_1} \right] -1}.
\end{equation}
The condition of reality of the above expression together with the
requirement of semi-classicality can be written as
\begin{equation}
0 \leq  \frac{1-2(\Re a_2)^2}{4\Re a_1} \ll \ln 2.
\end{equation}
Because $\Re a_1>0$, the two conditions must hold:
\begin{eqnarray}
(\Re a_2)^2 && \leq \frac{1}{2}, \\
\Re a_1 && \gg \frac{1-2(\Re a_2)^2}{4 \ln 2}.
\end{eqnarray}

For the special case $\Re a_2 = 0$ $(\alpha = 0)$ we obtain
\begin{equation}
\Re a_1 \gg \frac{1}{4 \ln 2},
\end{equation}
which yields the following constraint on dispersion
\begin{equation}
\sigma_Q \ll \sqrt{\ln 2} \approx 0.83.
\end{equation}

\subsection{Schr\"odinger cat composed of standard coherent states}

\begin{figure}[ht!]
\centering
\includegraphics[width=7cm,angle=0]{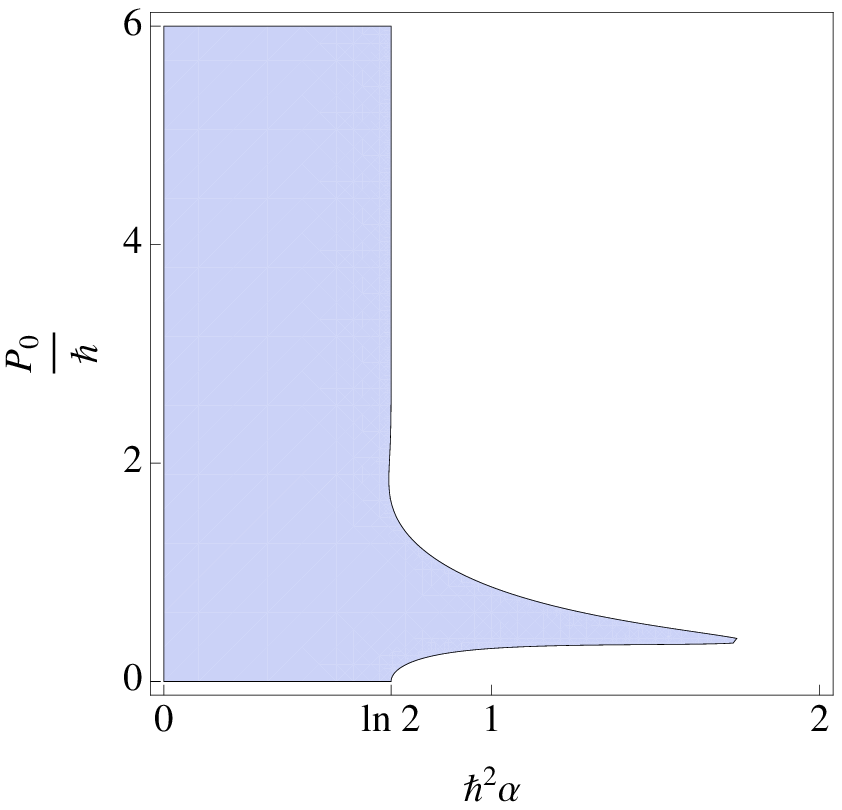}
\caption{For the state to be semiclassical  the parameters must
lie within the shadowed region. Semi-classicality  holds in a
neighborhood of the origin.} \label{SemiRegion2}
\end{figure}

For the Schr{\"o}dinger's cat state we have
\begin{eqnarray}
&&\int_{-\infty}^{+\infty} e^{X} |\Psi(X)|^2 dX = \nonumber \\
&&=N^2 e^{\frac{\hslash^2 \alpha}{2}}\left[1+e^{-2P_0^2 \alpha} \cos(P_0 \alpha \hslash)  \right] , \\
&&\int_{-\infty}^{+\infty} e^{2X} |\Psi(X)|^2 dX  = \nonumber \\
&&= N^2 e^{2\hslash^2 \alpha}\left[1+e^{-2P_0^2 \alpha} \cos(4 P_0 \alpha \hslash)  \right].
\nonumber
\end{eqnarray}
Based on the above we find
\begin{equation}
\lim_{\epsilon \rightarrow 0} \frac{\sigma_{q}}{\langle \hat{q} \rangle} =
\sqrt{\frac{e^{\hslash^2 \alpha}}{N^2} \frac{\left[1+e^{-2P_0^2 \alpha}
\cos(4 P_0 \alpha \hslash)\right]}{
\left[1+e^{-2P_0^2 \alpha} \cos(P_0 \alpha \hslash)  \right]^2} -1}.
\label{LimRelFlucSCS}
\end{equation}
In the limit $P_0 \rightarrow 0$, this equation simplifies to
\begin{equation}
\lim_{\epsilon \rightarrow 0} \frac{\sigma_{q}}{\langle \hat{q} \rangle} =
\sqrt{e^{\alpha\hslash^2}-1}.
\end{equation}
The semi-classicality condition  (\ref{semi2}) applied to (\ref{LimRelFlucSCS}) leads to the
following constraint
\begin{equation}
\frac{e^{\hslash^2 \alpha}}{N^2} \frac{\left[1+e^{-2P_0^2 \alpha} \cos(4 P_0
\alpha \hslash)\right]}{
\left[1+e^{-2P_0^2 \alpha} \cos(P_0 \alpha \hslash)  \right]^2}  \ll 2.
\label{SCatSem}
\end{equation}
In Fig. \ref{SemiRegion2} we show the region of the parameter
space for which $\lim_{\epsilon \rightarrow 0}
\frac{\sigma_{q}}{\langle \hat{q}\rangle} < 1$, indicating the
part of the parameter space within which the semi-classicality
condition is fulfilled.

In the special case when $P_0 \rightarrow 0$ (two peaks overlap) the condition (\ref{SCatSem}) simplifies to
\begin{equation}
\hslash^2 \alpha \ll \ln 2 \approx 0.693.
\end{equation}
By using (\ref{sigmaQCat}), the above constrain leads to
\begin{equation}
\sigma_Q \ll \sqrt{\ln 2} \approx 0.83,
\end{equation}
which, as expected, is in agreement with the case of the standard coherent state for $\Re a_2 =0$.

\subsection{Soliton-like state}

For the soliton-like state, the following integrals can be
determined:
\begin{eqnarray}
\int_{-\infty}^{+\infty} e^{X} |\Psi(X)|^2 dX &=& \frac{a \hbar}{ \sin(a\hbar)},\\
\int_{-\infty}^{+\infty} e^{2X} |\Psi(X)|^2 dX &=& \frac{2 a \hbar}{ \sin(2 a\hbar)} ,
\end{eqnarray}
for $a\hbar  < \pi$ in the first case and $a\hbar  < \frac{\pi}{2}$ in the second case.
The integrals are not convergent above these ranges.

By applying the above integrals to expression (\ref{LimRelFluc}), we obtain
\begin{equation}
\lim_{\epsilon \rightarrow 0} \frac{\sigma_{q}}{\langle \hat{q} \rangle} =
\sqrt{ \frac{\tan(a \hslash)}{a \hslash}-1}
\end{equation}
Employing the condition of semi-classicality (\ref{semi2}) we obtain
\begin{equation}
\tan(a \hslash) \ll 2 a \hslash,
\end{equation}
which can be solved numerically leading to
\begin{equation}
a \hslash \ll 1.17.
\label{a117}
\end{equation}

\subsection{Schr\"odinger cat composed of soliton-like states}

\begin{figure}[ht!]
\centering
\includegraphics[width=7cm,angle=0]{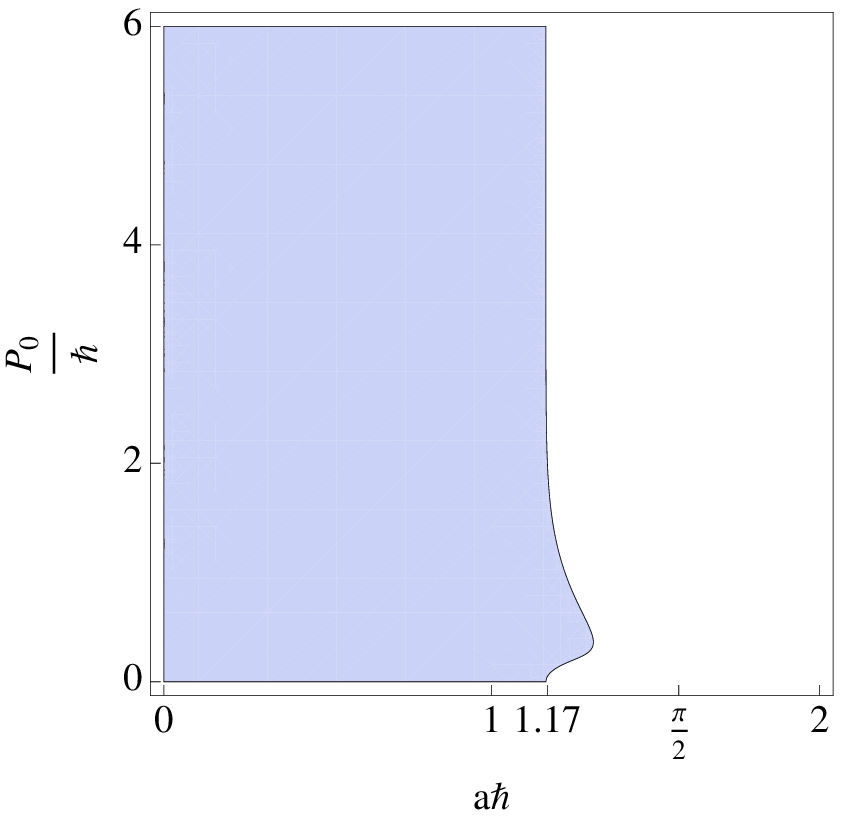}
\caption{For the state be semiclassical  the parameters must lie
within the shadowed region. Semi-classicality  holds in a
neighborhood of the origin.} \label{SemiRegion3}
\end{figure}

For the Schr\"odinger cat composed of soliton-like states on can
also find analytical formula for  $\lim_{\epsilon \rightarrow 0}
\frac{\sigma_{q}}{\langle \hat{q}\rangle}$. However, because the
obtained formula is quite long, we present, in Fig.
\ref{SemiRegion3}, only the the resulting constraint
$\lim_{\epsilon \rightarrow 0} \frac{\sigma_{q}}{\langle
\hat{q}\rangle} < 1$.
As expected, for $P_0 =0 $ the constraint (\ref{a117}) is recovered.

\section{Conclusions}

In this paper we have studied dynamics of the quantum bouncing
universe for a variety of quantum states.  We examined properties
of the following states: boxcar state, standard coherent state,
and soliton-like state, as well as Schr{\"o}dinger's cat states
constructed from each of them respectively. Characteristics of
these states such as quantum moments and Wigner functions were
investigated.

We have found a canonical transformation $\Gamma_1 \ni (q,p)
\rightarrow (Q,P) \in \Gamma_2$, which transforms initial
non-polynomial Hamiltonian $H_1= p \sin q$ into a simple linear
Hamiltonian $H_2=P$, which resembles the  Hamiltonian of a photon.
The unitary map $\mathcal{U}: \mathcal{H}_1 \rightarrow
\mathcal{H}_2$, corresponding to the canonical transformation was
also constructed.

Analysis of the Wigner functions for  considered states enabled us
to relate quantum dynamics with the classical phase space
structure. It was shown that, the  boxcar state, standard coherent
state, and soliton-like state can reproduce all trajectories on
the phase space. However, Sch\"odinger cat states, which are less
classical than the above states, can follow only the trajectories
with $P=0$, since $\langle \hat{P} \rangle=0$ for these states. By
analyzing the classical limit (large volumes) we have found that
all considered states have proper semi-classical behavior for some
values of parameters. We have determined ranges of the parameters.
The obtained semi-classical states are complementary to those
found in Refs. \cite{Corichi:2011rt, Corichi:2011sd}.

Furthermore, analysis of the Wigner functions for the considered
states leads  to the two general observations:
\begin{itemize}
\item  For the investigated states we have found that $W\left(
\langle X \rangle,\langle P \rangle \right)= 1/\hslash \pi.$ This
may suggest that such relation is fulfilled for some broad class
of pure states. For mixed states this relation does not hold as
can be seen for the decohered   Schr{\"o}dinger cat state. \item
We have found that the minimal area of the domain of the negative
values of the Wigner function may be bounded from below by the
factor $h/4$ for some class of states.  In other words, we
conjecture that $\oint  PdQ  \geq h/4$, for a region of the
negative values of the Wigner function. The approximate value
$h/4$ was found numerically for the Schr{\"o}dinger cat states
composed of two standard coherent states as well as for the
Schr{\"o}dinger cat state composed of two soliton-like states. For
the peculiar case of  the Schr\"odinger cat of two boxcar states
the above inequality is not fulfilled, which may result from the
fact that the uncertainty relation is not well defined for this
state. Therefore, it is possible that the relation $\oint PdQ
\geq h/4$ is valid only if the Schr{\"o}dinger-Robertson
uncertainty relation is satisfied.

\end{itemize}

Wigner's function can be reconstructed experimentally by applying
methods of quantum tomography. It is interesting to ask if such a
method could be applied also in the cosmological realm? The answer
to this question is however closely related to the meaning of
reduction of state in quantum cosmology. In laboratory, in order
to determine the shape of the Wigner distribution, an ensemble of
systems prepared in the same quantum state is necessary. Then, by
collecting measurements performed on each system, the structure of
the phase space distribution can be recovered by virtue of the
inverse Radon transform. Possibility of such a reconstruction in
cosmology is however uncertain due to an absence  of a sound
notion of quantum measurement  for the universe. The problem is
due to the fact that there is only one Universe. Therefore, one
cannot perform measurements on ensemble of identical universes.
However, measurements of expansion rate at different locations
(subsystems) can be done.  The average expansion at different
places is described by the same equation, with  similar initial
conditions. In such a case, statistical interpretation of quantum
mechanics could be applied, and extraction of the Wigner function
by quantum tomography would be, in principle, possible. This issue
deserves further investigations.

We have also shown that the notion of an entropy of squeezing,
introduced in our earlier paper \cite{Mielczarek:2012qs} is not
invariant under canonical transformations. Therefore, it  cannot
be used as a reliable intrinsic time parameter. The same applies
to quantum covariance, which only for some specific coordinates
evolves monotonically.

In this paper, we have   studied decoherence of the Schr\"odinger
cat state of the minisuperspace model, due to the interaction with
the environment. Such a decoherence is a toy model of what could
in fact happen in the very early universe. Firstly, this process
may describe decoherence of the triad orientation, as already
suggested in Ref.  \cite{Kiefer:2012kp}. This is a quite
attractive interpretation, since there are two equivalent triad
orientations allowed, and the process of decoherence affords an
explanation to the emergence of them in the classical world.

Secondly, the Schr\"odinger cat state can describe
superposition of two separated space points in the quantum
counterpart of the state of asymptotic silence (quantum BKL
scenario). One can speculate that, is such a phase, entire
Universe was in highly entangled state. Due to the the decoherence
process, the classical BKL scenario, characterized by existence of
almost independent homogeneous regions of space, emerged.

In the context of the {\it decoherence}, one can speculate what
follows: During the quantum phase the entire Universe was in a
highly entangled state with small quantum inhomogeneities
described by hypothetical {\it quantum} BKL scenario. Due to
expansion, the evolution entered {\it classical} BKL scenario
characterized by the existence of almost independent homogeneous
but anisotropic regions of  space. Later, each of these regions
turned into an isotropic region that could be modeled by the FRW
type spacetime. Due to entanglement during the quantum phase, all
space regions became correlated. Such quantum correlation may
explain observed causal connection of the cosmic microwave
background (CMB) radiation without need of inflation.

\acknowledgments

The work was supported by Polonium program N$^{\circ}$ 27690XK \textit{Gravity and Quantum Cosmology}.
Authors would like that to Karol {\.Z}yczkowski for helpful discussion.

\end{document}